\documentclass[]{spie}  
\usepackage[]{graphicx}
\usepackage{subfigure}
\usepackage{array}
\usepackage{url}
\usepackage{comment}

\def\pb{{\sc polarbear}}

\def\quad{{\sc quad}}

\def\arcmin{\hbox{$^\prime$}}

\newcommand{\beq}{\begin{equation}}
\newcommand{\eeq}{\end{equation}}
\newcommand{\bea}{\begin{eqnarray}}
\newcommand{\eea}{\end{eqnarray}}

\begin{document} 

\title{The POLARBEAR Experiment} 

\author{Zigmund D. Kermish\supit{a}, 
Peter Ade\supit{b}, 
Aubra Anthony\supit{c}, 
Kam Arnold\supit{a}, 
Darcy Barron\supit{d}, 
David Boettger\supit{d}, 
Julian Borrill\supit{e,f}, 
Scott Chapman\supit{n}
Yuji Chinone\supit{g}, 
Matt A. Dobbs\supit{h}, 
Josquin Errard\supit{i}, 
Giulio Fabbian\supit{i}, 
Daniel Flanigan\supit{a}, 
George Fuller\supit{d}, 
Adnan Ghribi\supit{a},
Will Grainger\supit{o}, 
Nils Halverson\supit{c}, 
Masaya Hasegawa\supit{g},
Kaori Hattori\supit{g}, 
Masashi Hazumi\supit{g}, 
William L. Holzapfel\supit{a},
Jacob Howard\supit{a}, 
Peter Hyland\supit{m}, 
Andrew Jaffe\supit{j}, 
Brian Keating\supit{d}, 
Theodore Kisner\supit{e}, 
Adrian T. Lee\supit{a}, 
Maude Le Jeune\supit{i}, 
Eric Linder\supit{k}, 
Marius Lungu\supit{a}, 
Frederick Matsuda\supit{d}, 
Tomotake Matsumura\supit{g}, 
Xiaofan Meng\supit{a},
Nathan J. Miller\supit{d}, 
Hideki Morii\supit{g}, 
Stephanie Moyerman\supit{d}, 
Mike J. Myers\supit{a}, 
Haruki Nishino\supit{a}, 
Hans Paar\supit{d}, 
Erin Quealy\supit{a}, 
Christian L. Reichardt\supit{a}, 
Paul. L Richards\supit{a,f},
Colin Ross\supit{n}, 
Akie Shimizu\supit{g}, 
Meir Shimon\supit{d}, 
Chase Shimmin\supit{a}, 
Mike Sholl\supit{k}, 
Praween Siritanasak\supit{d}, 
Helmuth Spieler\supit{k}, 
Nathan Stebor\supit{d}, 
Bryan Steinbach\supit{a}, 
Radek Stompor\supit{i}, 
Aritoki Suzuki\supit{a}, 
Takayuki Tomaru\supit{g}, 
Carole Tucker\supit{b}, 
Oliver Zahn\supit{k,l}
\skiplinehalf
\supit{a}Physics Department, University of California, Berkeley; \\
\supit{b}School of Physics and Astronomy, University of Cardiff; \\
\supit{c}Department of Astrophysical and Planetary Science, University of Colorado; \\
\supit{d}Physics Department, University of California, San Diego; \\
\supit{e}Computational Cosmology Center, Lawrence Berkeley National Laboratory; \\
\supit{f}Space Sciences Laboratory, University of California, Berkeley; \\
\supit{g}High Energy Accelerator Research Organization (KEK), Japan; \\
\supit{h}Physics Department, McGill University; \\
\supit{i}Laboratoire Astroparticule et Cosmologie (APC), Universite Paris 7; \\
\supit{j}Department of Physics, Imperial College; \\
\supit{k}Physics Division, Lawrence Berkeley National Lab; \\
\supit{l}Berkeley Center for Cosmological Physics (BCCP), University of California, Berkeley; \\
\supit{m}Physics Department, Austin College;\\
\supit{n}Physics Department, Dalhousie University; \\
\supit{o}Rutherford Appleton Laboratory, STFC; \\
}


\authorinfo{Author contact: zkermish@princeton.edu}

 
  \maketitle 

\begin{abstract}
We present the design and characterization of the {{\sc polarbear}} experiment.  {\sc polarbear} will measure the polarization of the cosmic microwave background (CMB) on angular scales ranging from the experiment's 3.5\arcmin \, beam size to several degrees.  The experiment utilizes a unique focal plane of 1,274 antenna-coupled, polarization sensitive TES bolometers cooled to 250 milliKelvin.  Employing this focal plane along with stringent control over systematic errors, \pb \,has the sensitivity to detect the expected small scale B-mode signal due to gravitational lensing and search for the large scale B-mode signal from inflationary gravitational waves.

{\sc polarbear} was assembled for an engineering run in the Inyo Mountains of California in 2010 and was deployed in late 2011 to the Atacama Desert in Chile.  An overview of the instrument is presented along with characterization results from observations in Chile.  
\end{abstract}



\section{INTRODUCTION}
\label{sec:intro}  

Incredible strides have been made in precisely characterizing the cosmic microwave background (CMB). 
  The wealth of information mined from the datasets of temperature and polarization anisotropies from many different experiments has led to our current understanding of the universe and the formation of a standard model of cosmology.  Observations of the CMB and other cosmological probes support the model of an expanding universe that originated in a hot Big-Bang, with structure formation driven through gravitational infall into initial adiabatic density perturbations.

While the model has survived significant observational scrutiny resulting in high precision measurements of many cosmological parameters, there remain large gaps in our ability to explain some of the fundamental aspects of this standard model.  Inflation is a proposed period of exponential expansion in the very early universe that explains several observables in our universe, such as the overall isotropy of the CMB and the observed flatness of the universe, and also provides a mechanism to generate the primordial density perturbations that serve as initial conditions for structure formation \cite{PhysRevD.23.347,PhysRevLett.48.1220,Linde:1981mu}.  However, inflationary theory itself remains relatively unconstrained by observations and it is therefore poorly understood.  A large parameter space of models exists that can adequately explain current observations, making it difficult to link the kinematic theory to fundamental physics without new measurements.

This proceeding presents the {\sc polarbear} experiment, one of several current experiments aiming to shed light on the nature of inflation.  The {\sc polarbear} instrument was first assembled for an engineering run in the Inyo Mountains of California in 2010 and was deployed in late 2011 to the Atacama Desert in Chile.  The instrument was sucessfully integrated and operated for both deployments, with observations in Chile currently ongoing.  {\sc polarbear} utilizes a unique 1274 element bolometric receiver to increase overall instrument sensitivity compared to the previous generation of high resolution instruments.  The experiment was designed to characterize the polarization of the CMB and measure the as-yet undetected B-mode polarization component across a wide range of angular scales.  The B-mode polarization on large angular scales carries a unique signature of inflation.  Detecting this signal would provide the proverbial `smoking-gun' of the inflationary model.  On smaller angular scales, the B-mode polarization is a probe of large scale structure due to gravitational lensing.  {\sc polarbear}'s sensitivity and angular resolution will allow it to also characterize the B-mode polarization on these small angular scales, providing measures of the sum of the neutrino masses and the equation of state of dark energy in the early universe.  

In Section \ref{sec:science}, we outline these scientific goals in more detail.  We present an overview of the instrument in Section \ref{sec:instrument}, and conclude with the current status of the instrument and some preliminary performance results in Section \ref{sec:performance}.  


\section{Science goals}
\label{sec:science}

A prediction of inflation is that both scalar and tensor perturbations are created.  Scalar perturbations are responsible for the density fluctuations that are observed in the primary temperature anisotropies and that serve as the seeds for structure formation.  They also produce the even parity ``E-mode" polarization that has been observed by several experiments, most recently the BICEP\cite{ 2010ApJ...711.1123C} and QUIET \cite{2011ApJ...741..111Q} experiments.  The level of tensor perturbations varies depending on the inflationary model, and the ratio of tensor to scalar perturbations, $r$, is currently constrained by temperature measurements alone to be $r < 0.22$ \cite{Komatsu:2009p1898}.  The tensor perturbations, or gravitational waves, imprint the CMB with odd-parity ``B-mode" polarization on large angular scales  \cite{PhysRevD.55.7368, PhysRevD.55.1830}.   The level of the B-mode signal produced by the inflationary gravitational waves provides a direct measure of $r$ and the energy scale of inflation.  An accurate characterization of the angular spectrum constrains the parameter space of inflationary models and even enables a reconstruction of the potential driving inflation in some simpler models \cite{Baumann:2008p419, Peiris:2006p1900}.  
  

While the tensor perturbations are the only \emph{primordial} source of B-mode polarization, there are other sources that need to be considered.   Gravitational lensing mixes E and B-modes, and thereby transforms a fraction of the initial E-mode signal into B-modes.  This leads to a B-mode polarization that peaks on small angular scales \cite{Lewis:2006p1816, Smith:2004p59}.  Lensing B-modes limit our ability to measure the inflationary B-mode signal for very small values of r\cite{Knox:2002p1732}. However, as a probe of large scale structure at z $\sim$ 2, these lensing B-modes will enable a rich range of science, including strong constraints on the neutrino masses and tests of general relativity and dark energy\cite{Smith:2008p1846}.

\begin{figure}[!t]
\begin{center}
\subfigure[]{
\raisebox{0.01in}{\includegraphics[width=3in]{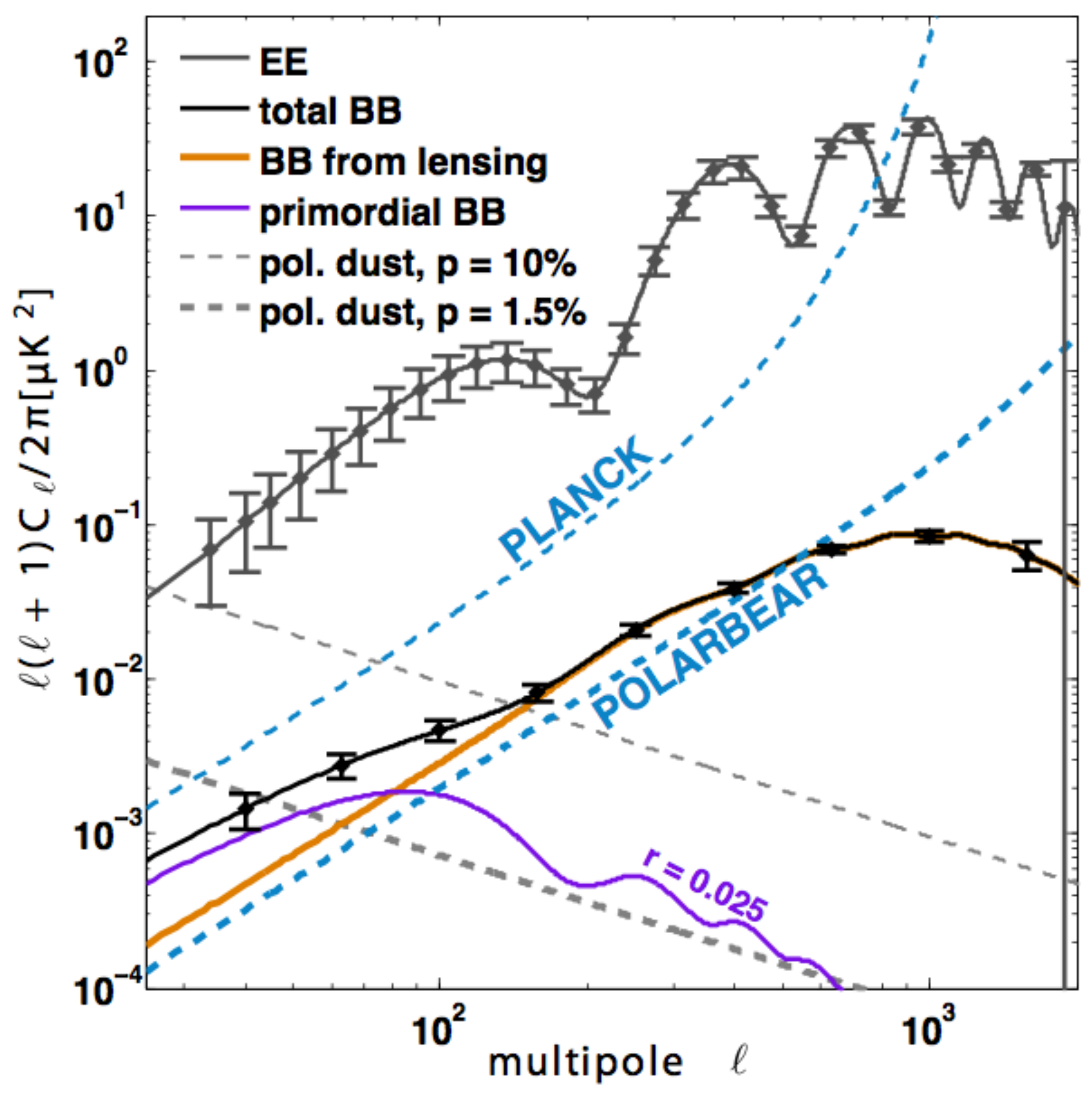}}
\label{fig:pb_error}
}
\qquad
\subfigure[]{
{\includegraphics[width = 3in]{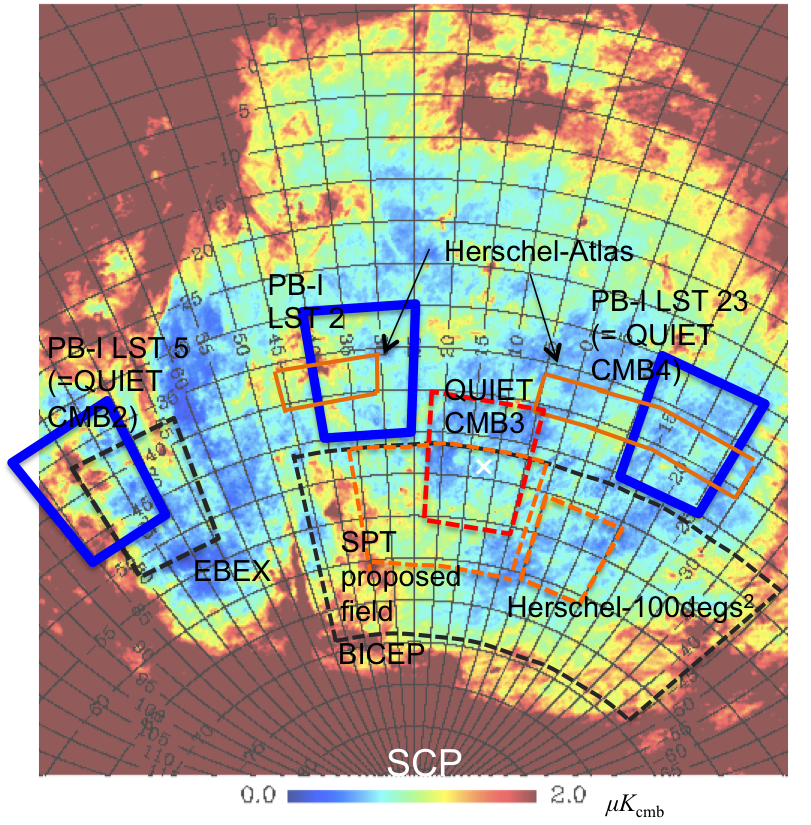}}
\label{fig:obspatch}
}
\caption{(a) Predicted error bars for the {\sc polarbear} experiment shown with theoretical E and B-mode polarization spectra assuming $r=0.025$.  The blue-dashed curves show the noise levels of {\sc polarbear} and PLANCK for comparison, both with $\Delta l = 30$ binning. (b) Proposed {\sc polarbear} observation patches shown outlined in blue along with theoretical galactic dust intensity.  Previously observed and proposed patches of observation from several complimentary experiments are also indicated.}
\label{}
\end{center}
\end{figure}

The {\sc polarbear} experiment was designed to characterize the B-mode polarization of the CMB on both large and small angular scales.  We detail the instrument design that enables this characterization in the next section.  The projected error bars on the E and B-mode polarization spectra are shown in Figure \ref{fig:pb_error}, along with the predicted instrument noise and a comparison to the Planck mission.  Error bars and the noise curve for {\sc polarbear} are calculated using the experiment design parameters summarized in Table \ref{tab:pb_sens}, assuming no foreground contamination and sub-dominant systematic error contributions. 	Noise equivalent temperatures (NETs) are estimated from the bolometer design parameters and estimates of the receiver loading and Chile sky temperature.  Factors contributing to the overall observing efficiency estimate include 70\% detector yield,  50\% efficiency on our observation patches and 65\% instrument operation time.    

\begin{table}[htdp]
\begin{center}
\begin{tabular}{|c|c|}
\hline
$\mathrm{NET}_{det}$ & 480 $\mu \mathrm{K} \sqrt{s}$ \\
$N_{det}$ & 1274 \\ 
$f_{sky}$ & 1.6 \% \\
Observing Efficiency & 18 \% \\
Observing time & 2 years\\
\hline
Temperature map depth & 6.3 $\mu \mathrm{K} - \mathrm{arcminute}$ \\
Polarization map depth & 8.9 $\mu \mathrm{K} - \mathrm{arcminute}$ \\
Beam FWHM & 3.5$^\prime$ \\
\hline
\end{tabular}
\end{center}
\caption{{\sc polarbear} instrument specifications}
\label{tab:pb_sens}
\end{table}%

The observation patches chosen for {\sc polarbear} are shown in Figure \ref{fig:obspatch} along with the planned and observed patches of several other experiments and the estimated galactic dust intensity.  The three $15^\circ \times 15^\circ$ observation patches were chosen to minimize the level of the galactic dust foreground while maximizing availability.  {\sc polarbear} observes in a single spectral band centered at 148 GHz instrument.  Overlap with the observation patches of CMB experiments at other frequencies will enable foreground characterization and removal if required.  

{\sc polarbear}'s designed sensitivity and 3.5\arcmin\,resolution will enable a greater than $10\sigma$ detection of the small-scale B-mode signal due to gravitational lensing.  When combined with data from the Planck satellite mission, data from {\sc polarbear} will yield a $1 \sigma$ error of 75 meV on the sum of the neutrino masses.  Even when not combined with the larger sky coverage of Planck, {\sc polarbear} will still provide a $1 \sigma$ error of 150 meV independently.  On large angular scales,  {\sc polarbear} will enable a 2$\sigma$ detection of $r=0.025$.  With this sensitivity, a significant majority of large field slow-roll models of inflation can be either detected or ruled out \cite{Pagano:2008jl}.  

\section{Instrument overview}
\label{sec:instrument}

A measurement of the B-mode polarization requires significant new developments in receiver technology over previous experiments.  
To characterize the B-mode polarization, the latest generation of millimeter wave receivers must be designed with both sensitivity and systematic error control in mind.  Since the B-mode signal is at least an order of magnitude below the observed E-modes, even low levels of cross or instrumental polarization can easily become sources of dominating systematic errors.  The faint signal also requires careful attention to all aspects of experiment design, including desired instrument resolution and observation strategies.

State-of-the-art millimeter-wave detectors are limited only by the photon noise of the thermal background they observe.  To increase overall sensitivity of the receiver and advance the instrument's mapping speed, large arrays consisting of thousands of such background-limited detectors that optimally sample a telescope's throughput are being used.  
{\sc polarbear} utilizes a unique 1,274 element lenslet coupled focal plane integrated with a large field of view telescope and cold reimaging optics to reach this goal.  Low levels of systematic error contamination are achieved with well-matched 3.5\arcmin\, dual-polarized beams, an optical design with low sidelobe response, and several levels of baffling to mitigate ground pickup.  Control and characterization of residual systematic errors are achieved by employing a stepped half-wave plate along with a scan strategy that leverages sky-rotation to provide polarization modulation.

\subsection{The Huan Tran Telescope}

\begin{figure}[htbp]
\begin{center}
\subfigure[]{
\includegraphics[width=3 in]{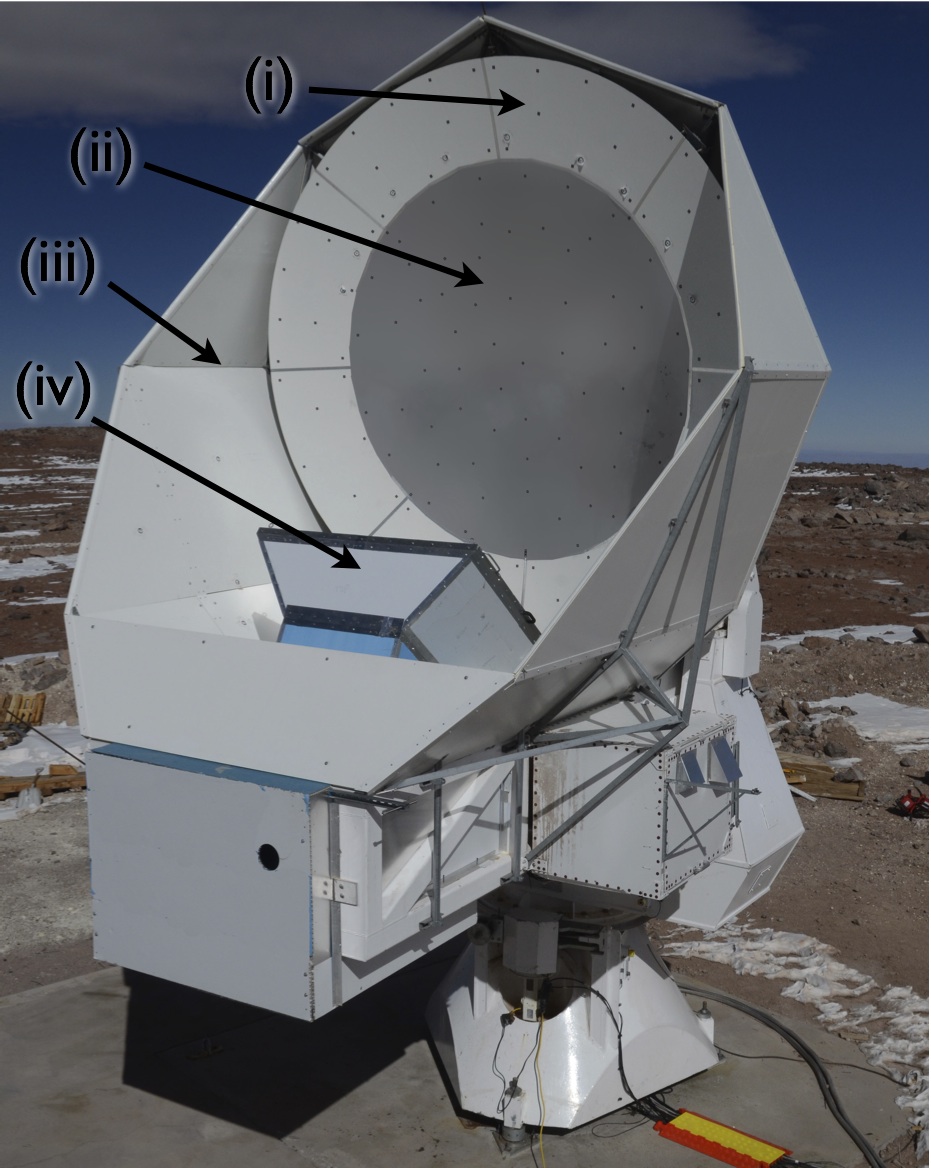}
\label{telescope_photo}
}
\qquad
\subfigure[]{
\raisebox{0.25in}{\includegraphics[width=3 in]{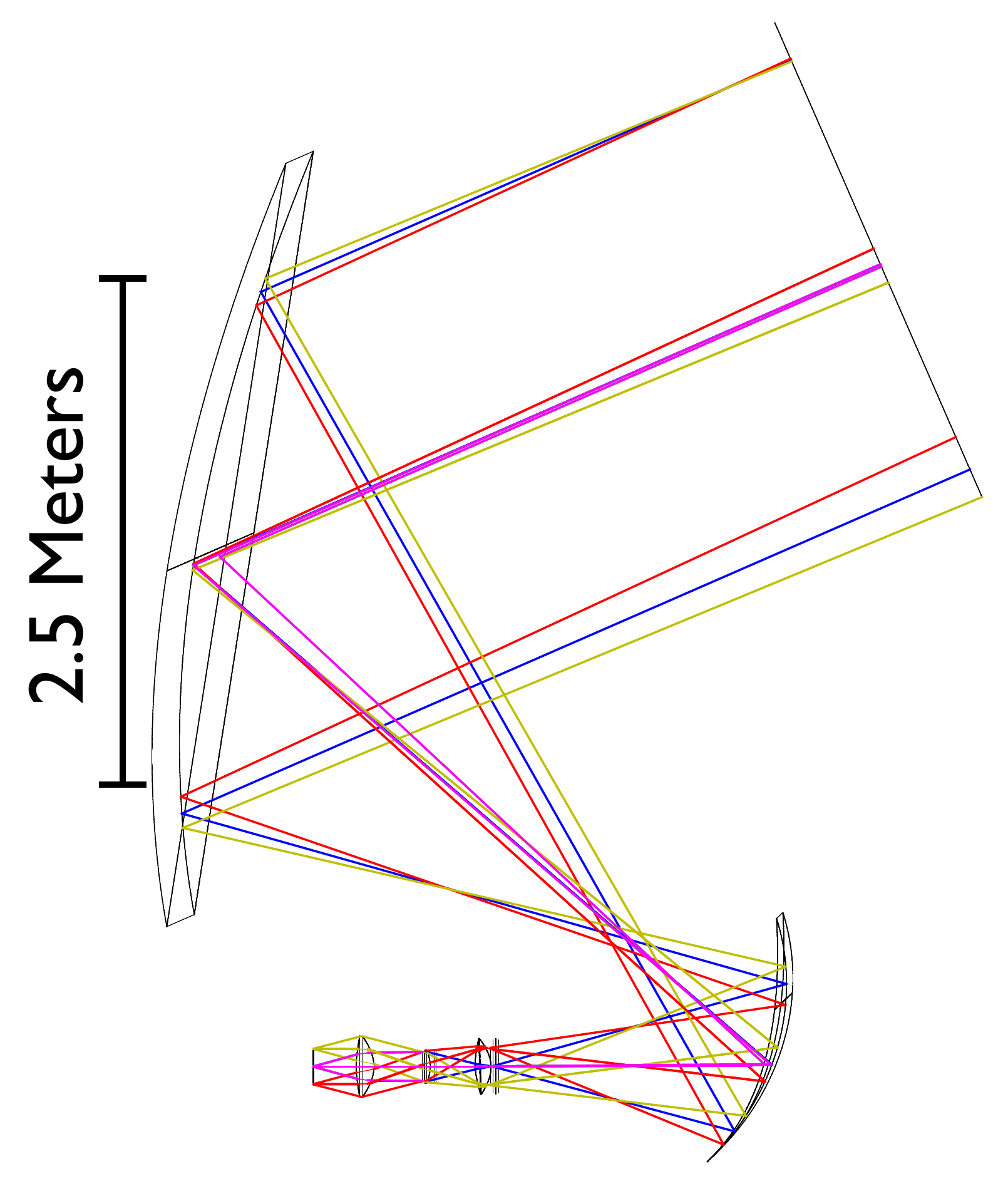}}
\label{optics_ray}
}
\caption{(a) The Huan Tran Telescope as assembled at the Cedar Flat site in the Inyo Mountains of California.  Indicated in the image are the (i) primary guard ring, (ii) precision primary mirror surface, (iii) co-moving shield, (iv), and prime-focus baffle.  The secondary mirror and receiver are not visible within their respective enclosures.  (b) A ray-tracing schematic of the telescope optics.  The focus created by the primary and secondary are reimaged by the cold re-imaging optics to a flat, tele-centric focal plane.}
\label{default}
\end{center}
\end{figure}

The Huan Tran Telescope (HTT) is an off-axis Gregorian design that satisfies the Mizuguchi-Dragone condition.   An advantage of off-axis telescopes is that they have clear apertures, lacking the secondary support structures obstructing the beam that are required for on-axis telescopes.  These structures can scatter or diffract signal from the ground into the main beam and severely limit performance.  The Mizuguchi-Dragone condition sets the tilt between the symmetry axes of the parent conics for the primary and secondary mirrors.  The resulting antenna has a rotationally symmetric equivalent paraboloid, giving low cross polarization and astigmatism over a large diffraction-limited field of view \cite{dragone, mizugutch}.  A Gregorian design has the advantage of a smaller secondary when compared to the alternative crossed design and allows easier baffling to prevent far sidelobes due to scattering at the receiver window.  One disadvantage is that the Gregorian design suffers from a smaller diffraction-limited field of view than the equivalent crossed-dragone design \cite{Tran:2008p1693}.  

A ray-tracing schematic of the HTT's optical design is shown in Figure \ref{optics_ray}.  HTT has a 2.5 meter primary mirror that is precision machined from a single piece of aluminum to a 53 $\mu m$ rms  surface accuracy, along with a lower-precision guard ring that extends the primary paraboloid out to 3.5 meters in diameter.  High surface accuracy and the monolithic design of the primary and secondary mirrors limit loss due to both diffuse scattering \cite{ruze} and knife-edge diffraction in the case of segmented mirrors.  The 2.5 meter primary gives a 3.5 arcminute beam at 150 GHz, allowing the experiment to probe out to $l \sim 2500$ and characterize the peak of the lensing B-mode spectrum at $l \sim1000$.  


\begin{table}[htdp]
\begin{center}
\begin{tabular}{| c | c | c |}
\hline                       
Description & Specification & Achieved performance \\
\hline
Maximum Az/El Velocity & $4^{\circ}/s$ & $4^{\circ}/s$ \\
Maximum Az/El Acceleration & $2^{\circ}/s^2$ & $2^{\circ}/s^2$ \\
Azimuth Travel & $\pm 200^\circ$ &  $\pm 200^\circ$ \\
Elevation Travel & $+40^\circ$ to $+90^\circ$ &  $+40^\circ$ to $+90^\circ$\\ 
Pointing reconstruction error & 10 arcsec & 12 arcsec rms \\
\hline
\end{tabular}
\end{center}
\caption{Telescope performance specifications and achieved performance}
\label{vertex:req}
\end{table}%

The telescope was built by VertexRSI\footnote{http://www.gdsatcom.com/vertexrsi.php} (now a part of General Dynamics).  The specifications given to Vertex for the telescope performance are summarized in Table \ref{vertex:req} along with the performance we've currently achieved.  A photo of the telescope fully assembled at the James Ax Observatory in Chile is shown in Figure \ref{telescope_photo}

\subsection{Cryogenic receiver}

\begin{figure}[htbp]
\begin{center}
\includegraphics[width=\textwidth]{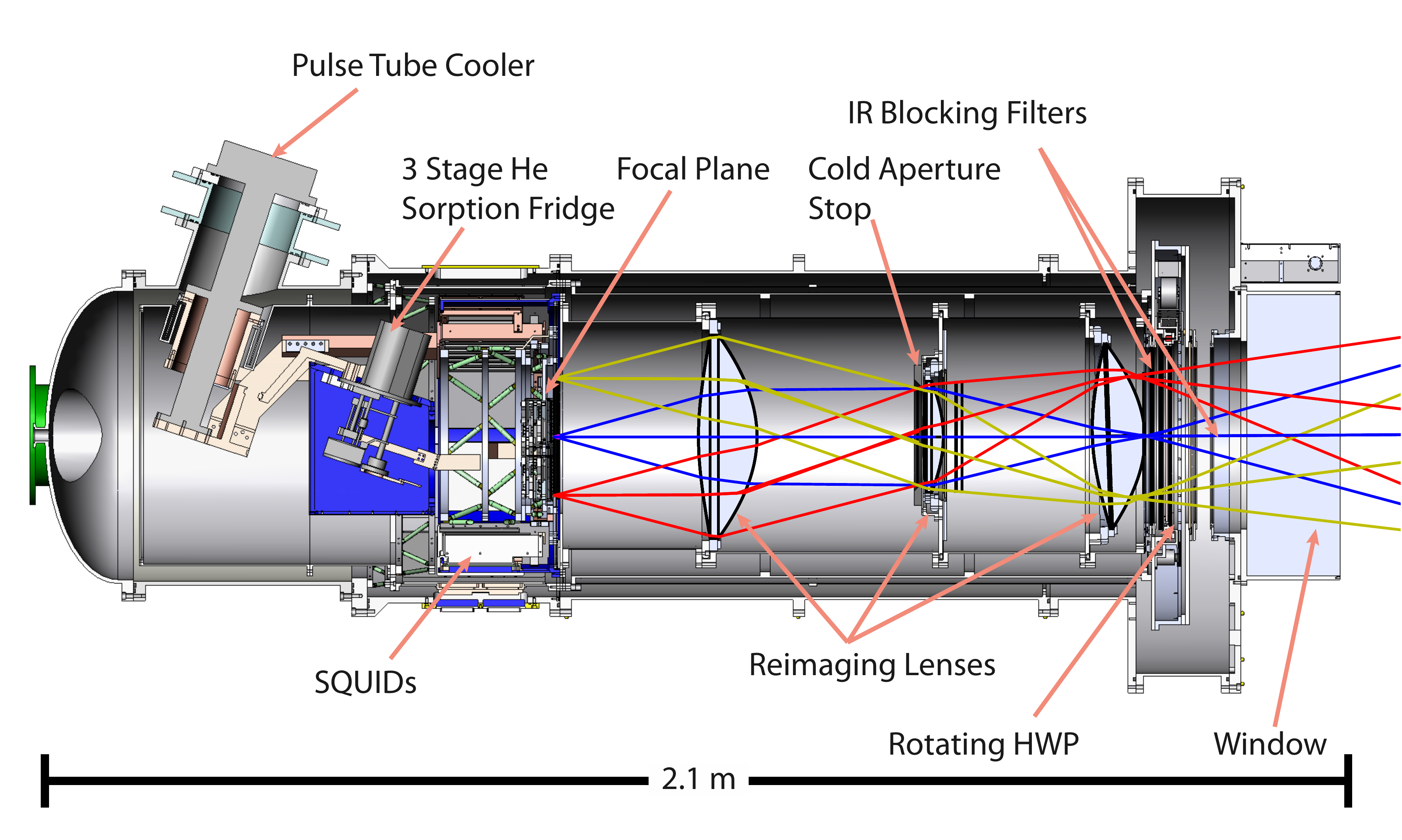}
\caption{A cross-section drawing of the {\sc polarbear} receiver.}
\label{receiver_overview}
\end{center}
\end{figure}

There are several overarching design goals for the {\sc polarbear} receiver:
\begin{itemize}
\item{Create a cryogenic environment that allows our focal plane to cool to $\sim$0.25 Kelvin with a $> 20$ hour hold-time.}
\item{Sufficiently cool the receiver and enclosed optical elements such that their in-band emission is lower than the expected in-band emission of the Chilean atmosphere.}
\item{Cool the superconducting quantum interference devices (SQUIDs) used for the multiplexed readout below their transition temperature to  $<$ 6 Kelvin}
\end{itemize}

All of these are effectively requirements to minimize the noise of the instrument.  Bringing the focal plane below 0.25 Kelvin makes the thermal carrier noise a subdominant term when compared with expected thermal background loading noise from the Chilean atmosphere.  Similarly, to keep the instrument truly background limited, the dewar itself must contribute a subdominant term to the overall thermal background noise when compared to the sky.  Cooling the SQUIDs is not only necessary for their functionality, but also to achieve sufficient transimpedance for high gain in the multiplexed readout. 

Cooling power for the {\sc polarbear} receiver is provided by two closed cycle refrigerators.  Our base temperatures of 50 Kelvin and 4 Kelvin are provided by a pulse tube refrigerator.  A commercial PT415 model from Cryomech, Inc.\footnote{http://www.cryomech.com/} is used.  The PT415 pulse tube used in the {\sc polarbear} receiver provides $\sim40$ Watts of cooling power at 45 Kelvin on the first stage and $\sim1.5$ Watts at 4.2 Kelvin on the 2nd stage.  

Subkelvin cooling is provided by a three stage, helium sorption fridge provided by Chase Research\footnote{{http://www.chasecryogenics.com/}}.  Cooling power is provided by pumping on condensed $^4$He and $^3$He in the various stages using charcoal pumps.  Our particular fridge uses a sacrificial $^4$He cycle which condenses off the 4 K mainplate to provide a $<$ 1 K condensation point for $^3$He used in the subsequent `ultrahead' and `interhead' stages.  The interhead stage simply acts as a buffer to intercept thermal loads from wiring and structural members, allowing the ultrahead to reach $\sim 0.25$ K.  Hold times of at least 24 hours are possible with loads of 80 $\mu$W on the interhead and up to 4 $\mu$W on the ultrahead giving operating temperatures of 375 mK and 250 mK, respectively.  Additional cooling power is also available at the heat exchanger, with $\sim$ 100 $\mu$W at an equilibrium temperature of $\sim$ 1.5 K.  

The refrigerators can be seen in the receiver cross-section in Figure \ref{receiver_overview}, along with various other components of the receiver.  The main components to note are: the 30 cm Zotefoam window; IR blocking thermal filters; the rotating half-wave plate, cooled to $\sim$80 Kelvin; reimaging lenses cooled to $\sim$6 Kelvin; the milliKelvin focal plane and the SQUID readout boards, cooled to $\sim$4 Kelvin.  The cryogenic environment achieved allows the focal plane to cool to 250 milliKelvin with a resulting hold time greater than 30 hours.

\subsubsection{Thermal filtering}

A challenge to deploying large-format arrays is in developing cryogenic systems with sufficiently large apertures for the high optical throughput.  The difficulty here is in rejecting radiation outside of the frequency band of interest to achieve a working cryogenic system while simultaneously mitigating in-band emission from the filter elements used.  A receiver with cold, low in-band emissivity optical elements is required to assure that the instrument sensitivity is limited by photon background loading from the sky rather than the receiver itself.  This task is further complicated by the need to limit systematic effects that can be introduced by filtering elements.  

This was addressed in {\sc polarbear} with a filter stack consisting of several hot-pressed multilayer low-pass filters and single layer IR shaders\cite{Ade:2006p359, Tucker:2006p358}, supplemented with an IR-absorbing 3 mm thick sheet of porous teflon.  These filters, combined with the receiver's Zotefoam vacuum window and cold re-imaging optics, mitigate the IR radiation incident on various cryogenic stages.  The result is a cryogenic environment that meets our design goals.  

\subsubsection{Polarization modulation}

\begin{figure}[!h]
\begin{center}
\includegraphics[width=4in]{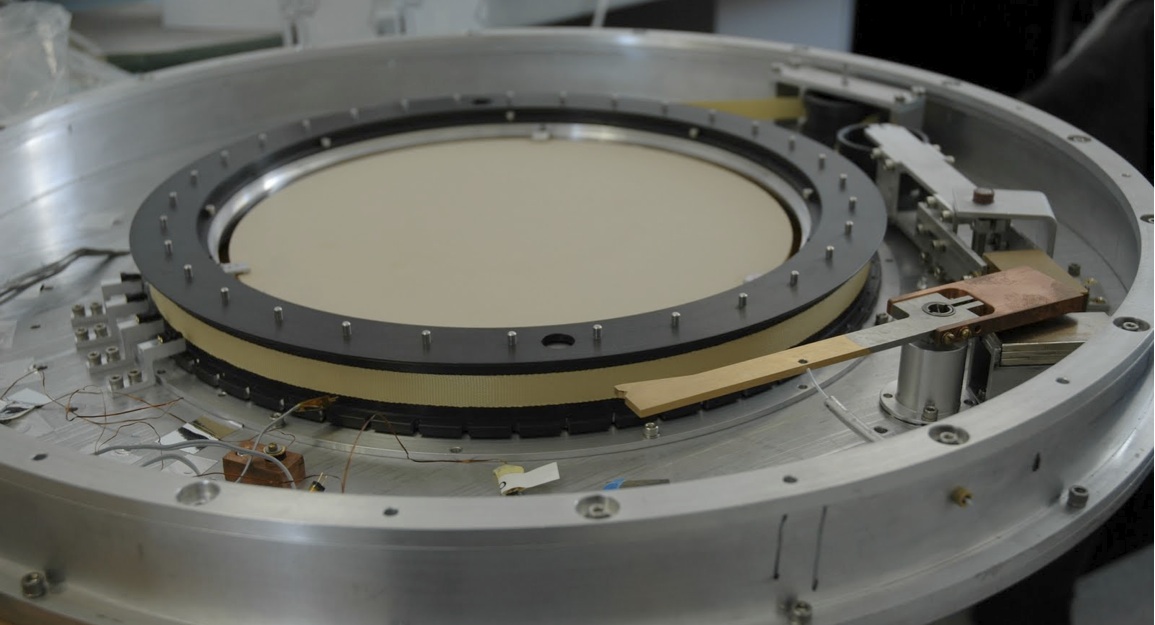}
\caption{A photo of the anti-reflection coated sapphire HWP and its rotation mechanism used in {\sc polarbear}.}
\label{fig:hwp}
\end{center}
\end{figure}

Modulation of the polarization on the sky with respect to the receiver and detectors helps mitigate systematic errors from instrument non-idealities, easing requirements on the levels of instrumental and cross-polarization and allowing such systematic errors to be better characterized and corrected \cite{Shimon:2008p512}.  Modulation occurs naturally with the sky rotation available from our mid-latitude observation site.  

Polarization modulation can also be introduced with optical elements, such as a rotating half wave plate (HWP).  {\sc polarbear} uses a single-crystal disk of A-plane cut sapphire with an anti-reflection coating of TMM as a HWP.  The HWP is cooled to $\sim$80 Kelvin to mitigate its contribution to detector loading.  A stepped rotation mechanism is employed to aid in mitigating systematic errors in conjunction with sky rotation in our scan strategy.  The anti-reflection coated HWP and its rotation mechanism are shown in Figure \ref{fig:hwp}.

\subsubsection{Cold reimaging optics}

The telescope focus lies just in front of the first reimaging lens of the receiver shown in Figure \ref{optics_ray}.  The field, aperture,  and collimating lenses serve to reimage the curved focus of the telescope to a flat, telecentric focal plane that can be coupled to planar, micro-fabricated detector arrays.  Along with the cold aperture stop, the reimaging optics were designed to give the telescope a $2.3^{\circ}$ diffraction-limited field of view for a 19cm focal plane.  The lenses are machined from ultra-high molecular weight polyethylene (UHMWPE) and cooled to $\sim$6 Kelvin to reduce their in-band emission.  Anti-reflection coatings consisting of a single layer of porous polytetrafluoroethylene (PTFE) are adhered to all lens surfaces using a vacuum hot-press process.    

\subsection{Focal plane architecture}

\begin{figure}[t!]
\begin{center}
\subfigure[]{
\raisebox{0.025in}{\includegraphics[width=3.25in]{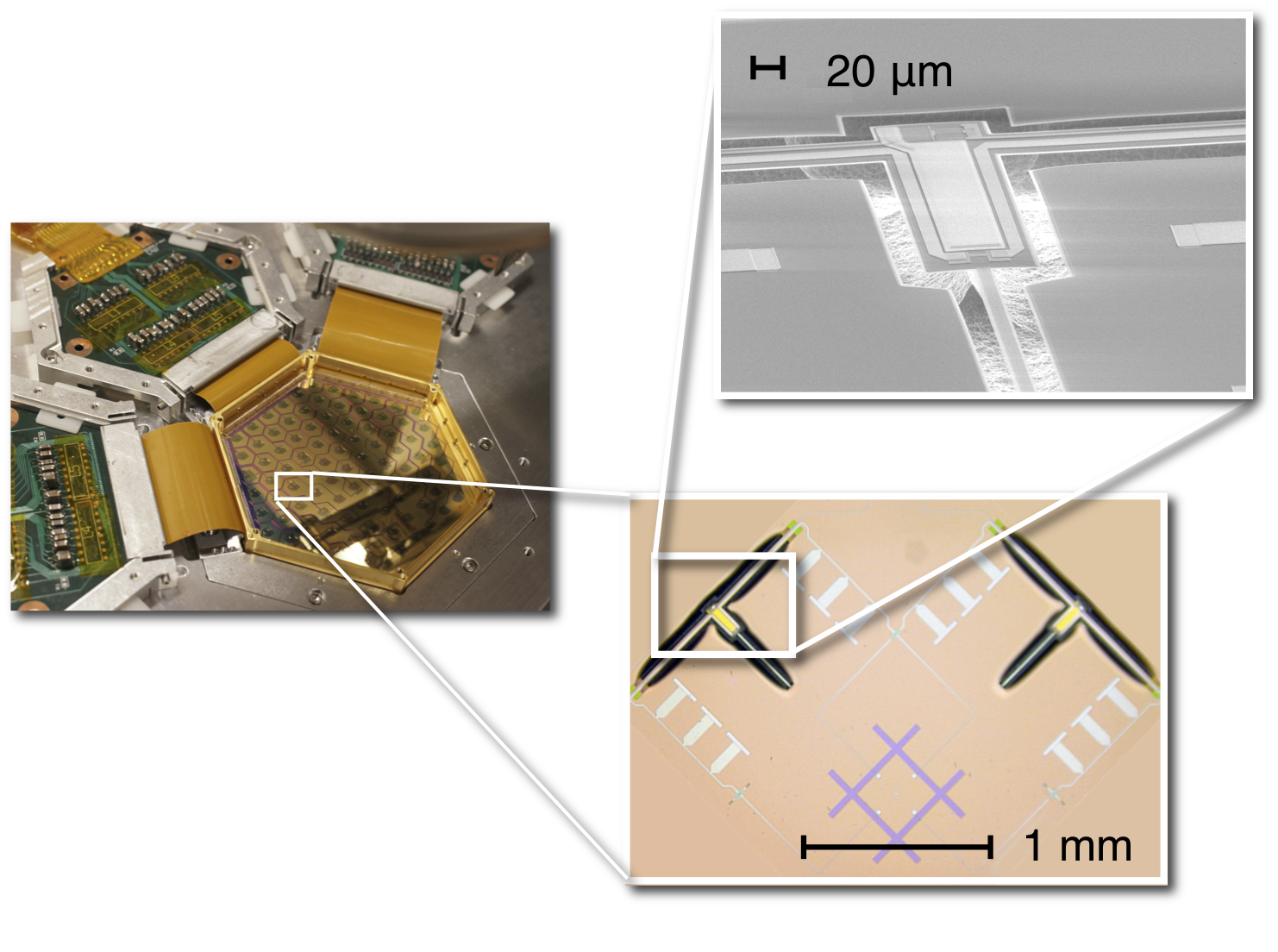}
}
\label{device_wafer}
}
\qquad
\subfigure[]{
\includegraphics[width=2.5 in]{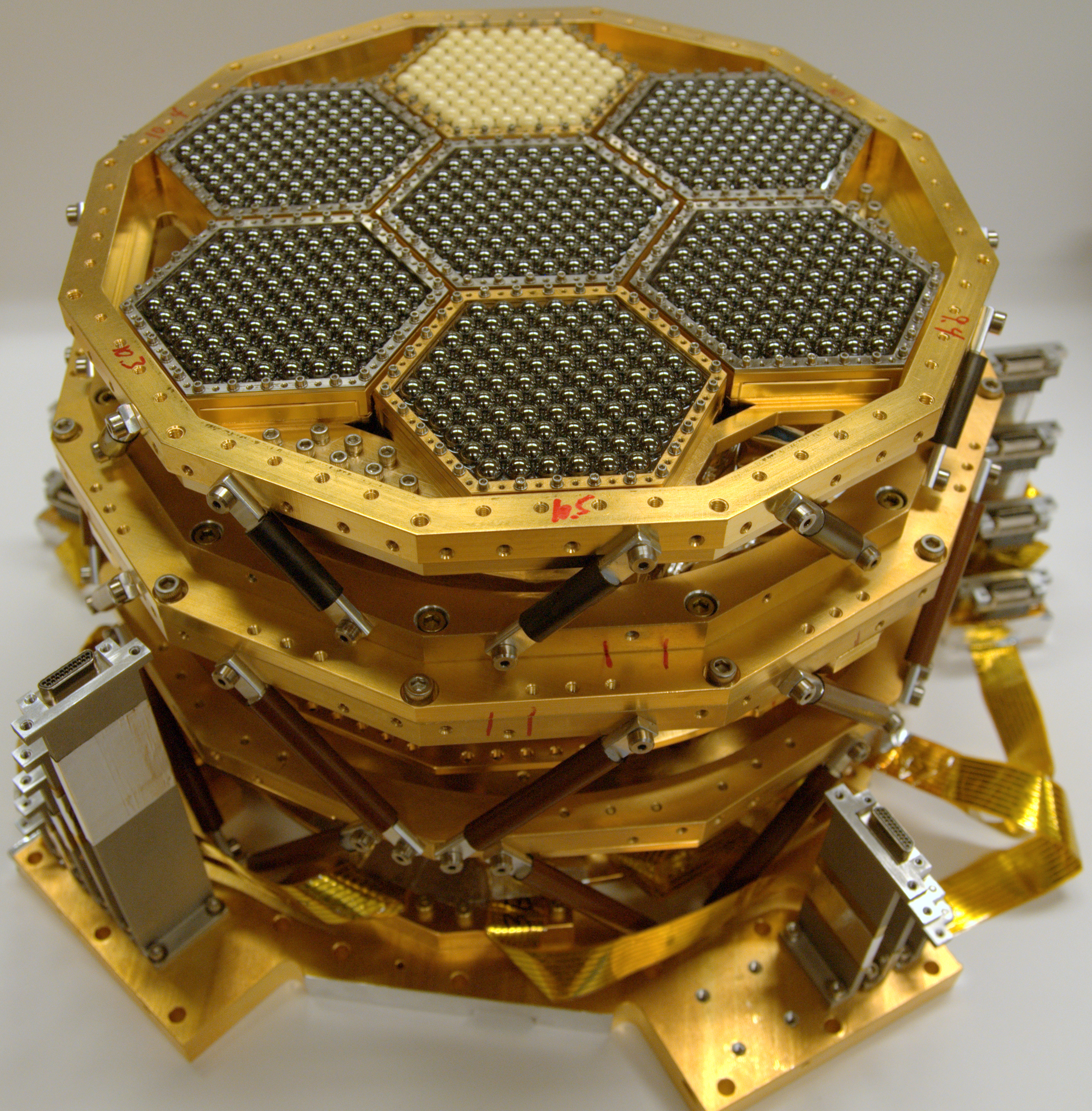}
\label{fp_assem_photo}
}
\caption{(a) A {\sc polarbear} device wafer.  The left image shows a device wafer installed in its wafer holder below a lenslet array.  The bottom right inset shows an image of a single pixel and the top right inset further magnifies, showing an SEM image of the bolometer island. (b) A photograph of the fully assembled focal plane, support structure and milliKelvin wiring.  Note that one module that is white in this photograph is using hemi-spherical lenselts that are made from alumina rather than silicon.  We initially used alumina lenslets since they are cheaper and more readily available than silicon and could be used as prototypes when we were first developing our fabrication techniques for the anti-reflection coatings. }
\label{fp_assem}
\end{center}
\end{figure}

The focal plane architecture of  {\sc polarbear} consists of antenna-coupled transition edge sensor (TES) detectors \cite{Myers:2005p1696}.  Radiation is coupled onto microstrips using polarization sensitive lenlset-coupled slot antennas \cite{Chattopadhyay:1998p2008}.  Silicon lenslets and a silicon spacer directly above the antennas provide a simulated elliptical lens, defining the beam directivity and narrowing the Gaussian beam width inherent to the antenna \cite{Zmuidzinas:1995p2006, Filipovic:1993p1801}.  On-chip spectral band filtering is achieved using resonant structures before power is dissipated at a load resistor located on a thermally isolated island.  A TES thermistor sits on this island and converts changes in optical power to modulated current through the AC biased thermistor.  Each pixel in a {\sc polarbear} module contains detectors for two orthogonal polarizations.  Figure \ref{device_wafer} shows the `readout side' of a single device wafer with 91 pixels, along with unfolded readout circuitry boards.  Details on the optimization, design and construction of these device wafers can be found elsewhere in these proceedings \cite{Kams_proceedings}.  The readout circuitry boards fold in to a compact module, lying within the footprint of the device wafer and allowing for a close-packed focal plane.  Seven wafer modules are used in the fully populated {\sc polarbear} focal plane for a total of 1274  detectors paired in 637 pixels.  The fully assembled focal plane insert deployed to Chile is shown in Figure \ref{fp_assem_photo}.

\subsection{Multiplexing readout}

A fundamental issue that one quickly runs into when developing a large detector array is in routing cryogenic wires.  Bias signals must be sent from room temperature electronics down to cryogenic focal planes and readout signals must get back out without exceeding the cryogenic budgets of feasible milliKelvin refrigerators.  For arrays of TES bolometers, this has been addressed by several multiplexing technologies.  Current through a low impedance, voltage-biased TES detector can be measured using a super-conducting quantum interference device (SQUID) by converting the current to magnetic flux with an inductor coil.  For {\sc polarbear} we use a frequency domain multiplexer\cite{Lanting:2005ep, Dobbs:2011uf}.  The multiplexer allows many detectors to be biased with different AC bias frequencies on a single pair of wires and readout on another, using a single SQUID per multiplexed set.  {\sc polarbear} was one of the first experiments deployed utilizing a newly developed digital frequency domain multiplexer (DfMux)\cite{Smecher:2012jm} with an 8$\times$ multiplexing factor.  


\section{Status and performance}
\label{sec:performance}

{\sc polarbear} was first deployed on the HTT in 2010 at the Cedar Flat site of the CARMA array in the Inyo Mountains of California.  The receiver was equipped with a subset of three of the seven planned detector arrays and only 1/3 of the readout for this engineering run.  The entire instrument was integrated and underwent several characterization tests.  Among the characterization tests performed were intensity maps of bright sources for beam characterization and telescope pointing evaluation and tests of polarization purity through measurements of ground based and celestial sources.  The successful engineering run help to refine the development of the receiver and full focal plane in preparation for our science deployment to Chile.  In late September of 2011, concrete was poured for the James Ax Observatory at an altitude of 5200 m on Cerro Toco in the Atacama desert of Chile.  In the 4 months that followed, the HTT was reassembled at the site and the {\sc polarbear} receiver integrated.  First light with the fully integrated experiment was achieved on January 10th, 2012 with a scan of Jupiter.  In this section, we present preliminary results on the instrument performance in Chile.    
  
\subsection{Beam maps}
\label{sec:beams}

Maps of planets such as Jupiter and Saturn provide a unique `all-in-one' method to measure several aspects of the instrument's performance.  When it subtends a solid angle significantly smaller than the designed beam, a planet provides a way to probe the structure of the beams themselves.  This allows the designed beam-size and ellipticity to be verified for every detector.  The location of each pixel on the sky relative to the telescope boresite pointing, known as the pixel offset, is also measured from these maps to allow co-adding of individual detector maps.  Gain-calibrating the two detectors of a single pixel via the response to atmospheric signal or using the beam map itself normalized to a known source temperature, the maps can also be used to probe the differential beam properties of each pixel by differencing the two orthogonal polarizations of each pixel and characterizing the residual multipole moments.  Beam maps also provide a means to measure both the fractional throughput bandwidth product, $\eta\Delta \nu$, and the detector NET directly for a known source temperature and angular extent.  

\begin{figure}[t!]
\begin{center}
\subfigure[]{
\includegraphics[width=3.in]{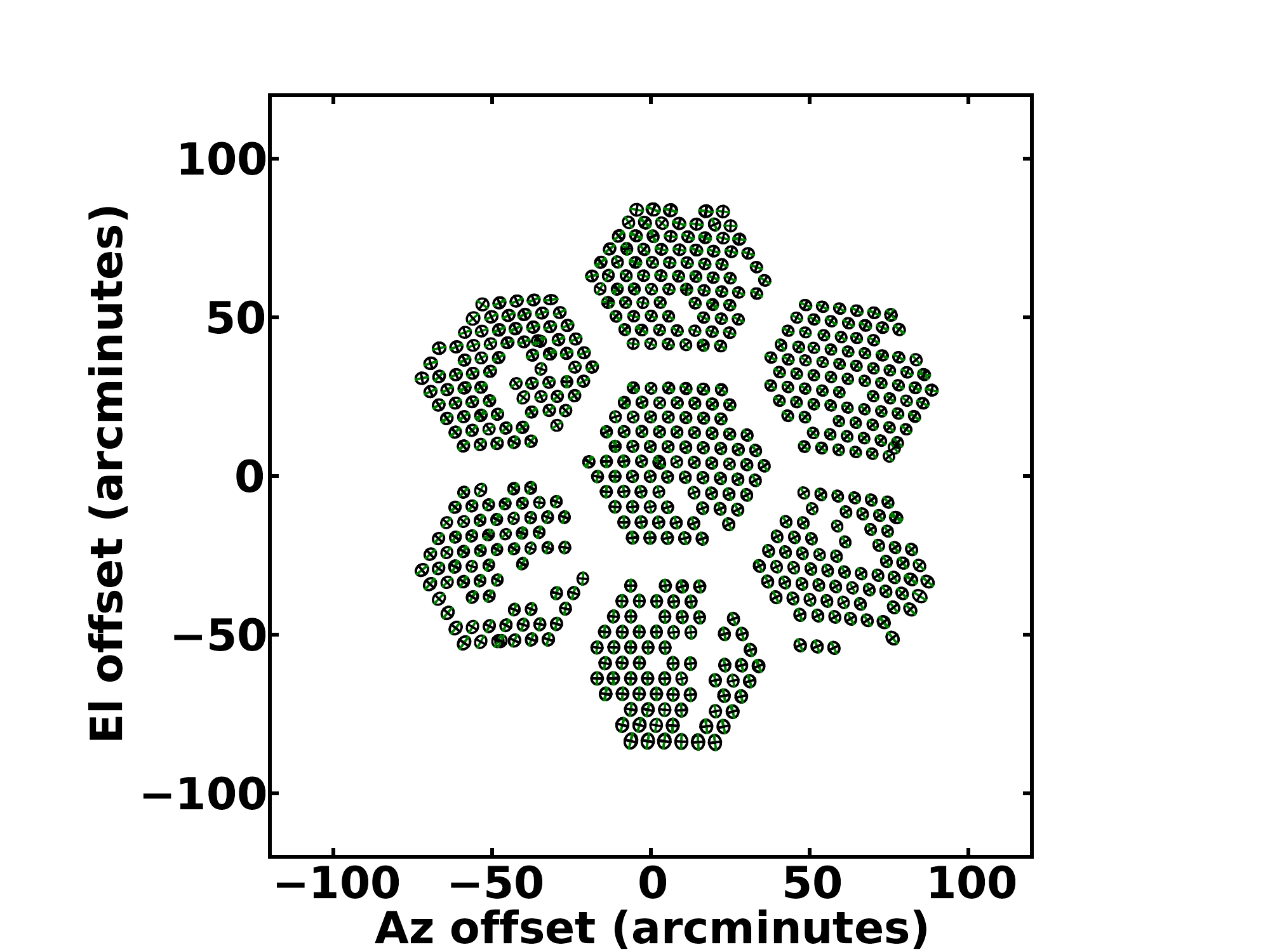}
\label{fig:beamprms}
}
\qquad
\subfigure[]{
\includegraphics[width=3.in]{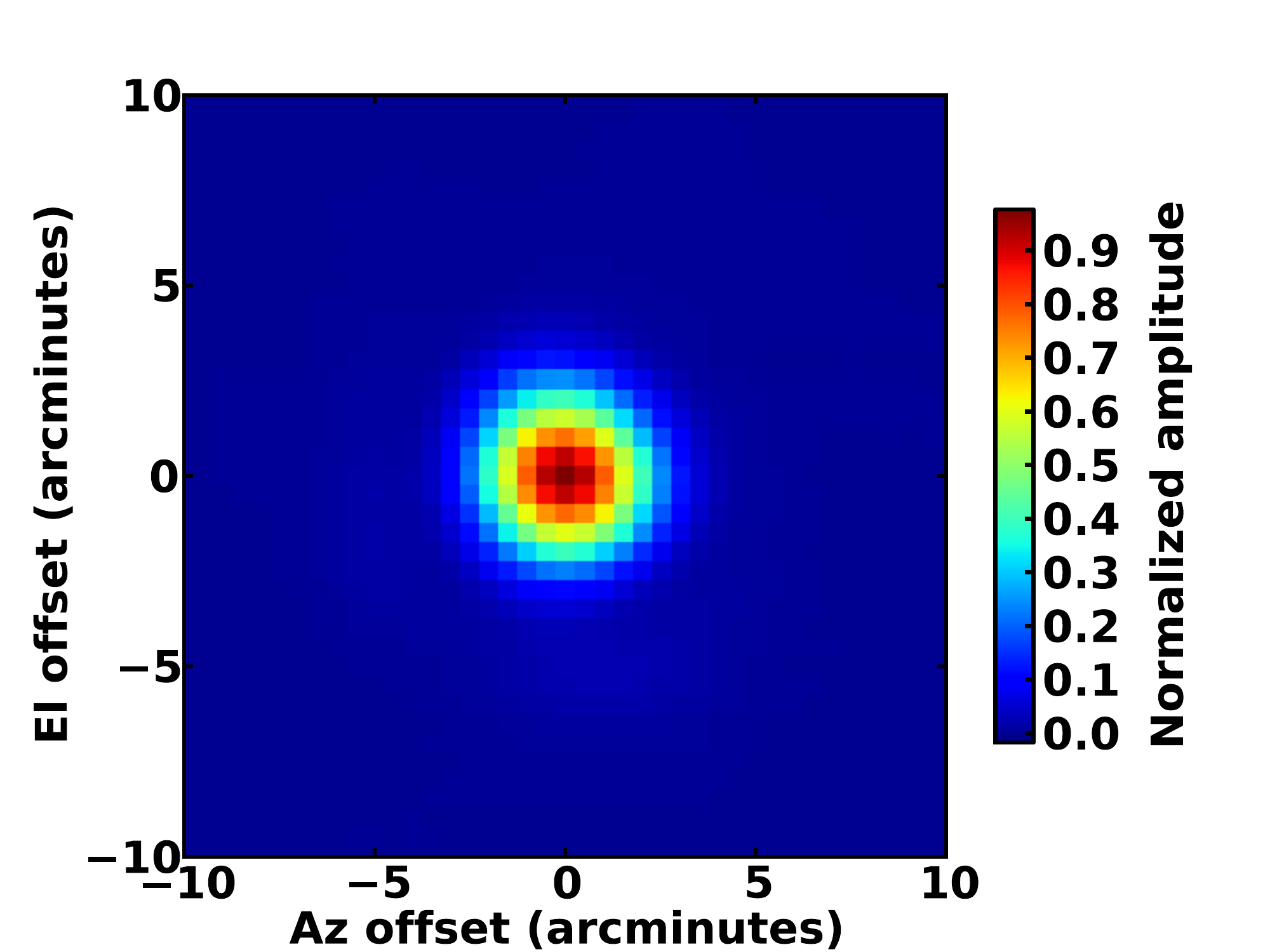}
\label{fig:sat0512coadd}
}
\caption{(a) Fit gaussian beam parameters (centroid and major/minor axes) for the 1015 active bolometer channels in the {\sc polarbear} focal plane from 18 observations of Saturn.  (b) The resulting co-added instrument beam from all detectors in 5 separate observations of Saturn.}
\label{}
\end{center}
\end{figure}

\begin{figure}[htdp]
\begin{center}
\subfigure[]{
\includegraphics[width=3.in]{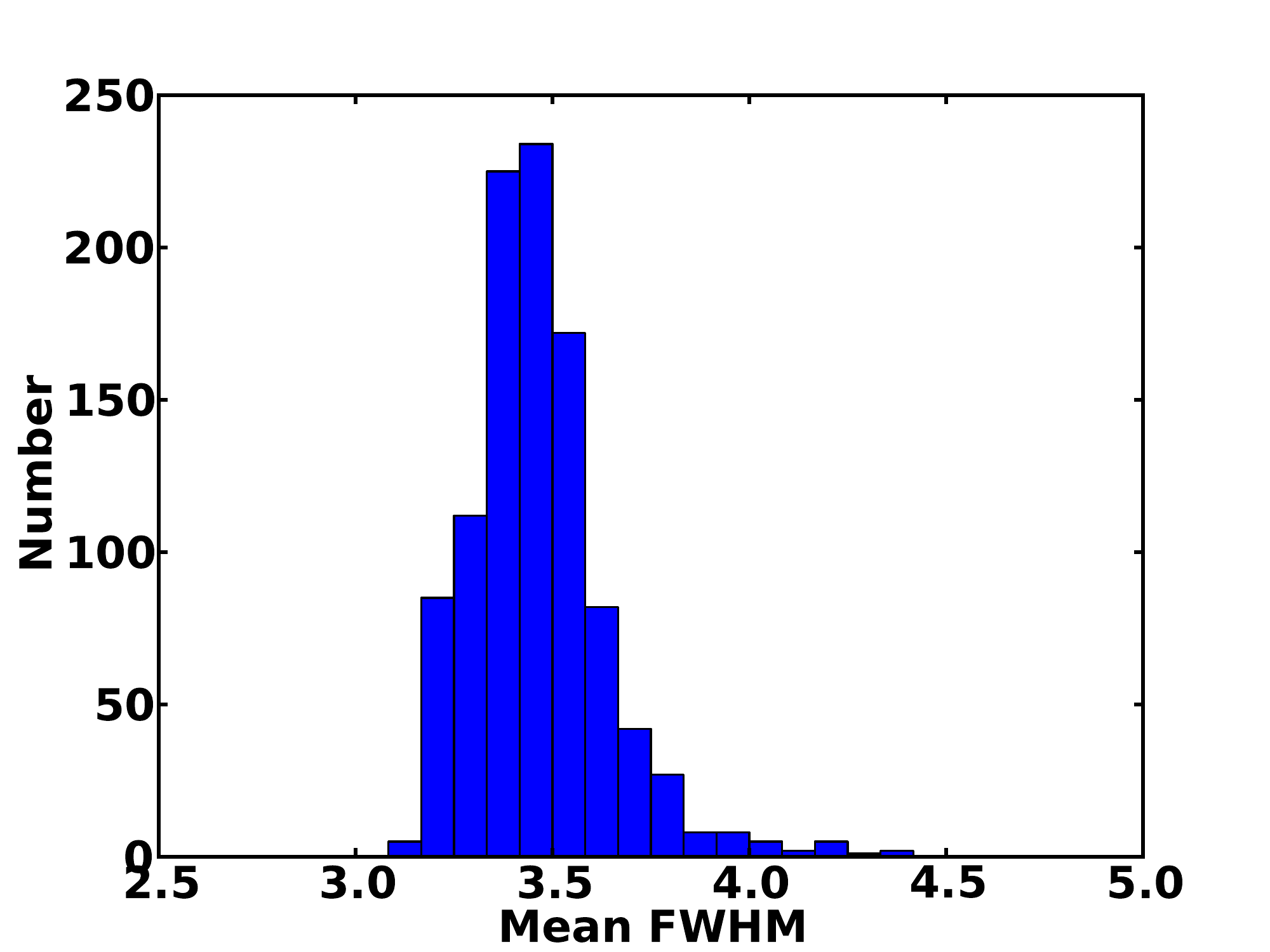}
\label{fig:fwhm}
}
\qquad
\subfigure[]{
\includegraphics[width=3.in]{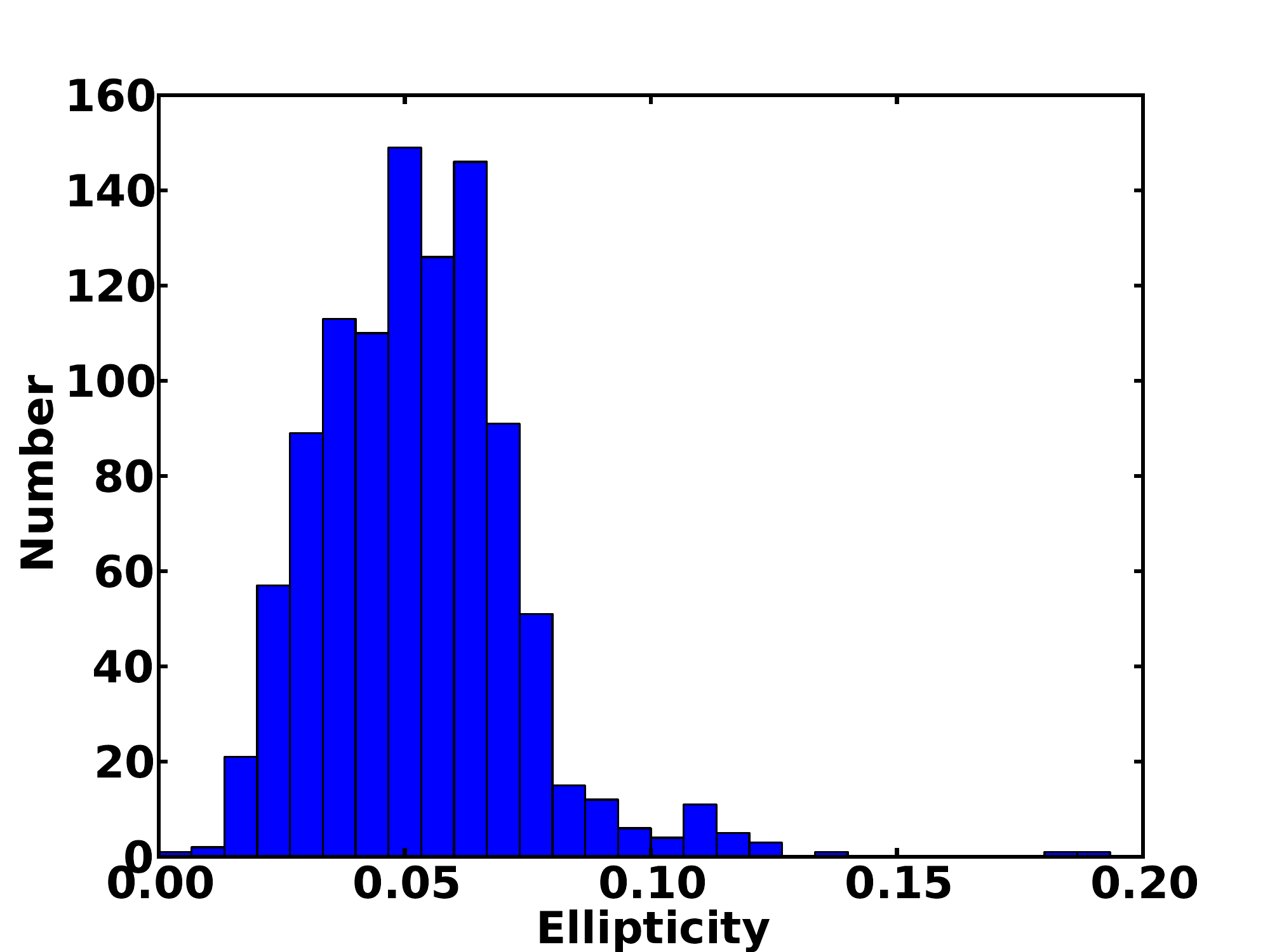}
\label{fig:ellip}
}
\caption{(a) Histogram of the fit full width at half maximum (FWHM) for all detector beams across the array.  (b) Histogram of the beam ellipticities for all detector beams across the array.}
\label{}
\end{center}
\end{figure}

Figure \ref{fig:beamprms} shows the fit beam parameters for the entire 7-wafer focal plane resulting from several observations of Saturn from Chile.  The current optical yield is 1015 out of 1274 possible detectors, or 80\%.  In Figure \ref{fig:sat0512coadd}, the beam offsets are used to coadd all detectors from five separate observations of Saturn and produce a high fidelity map of the overall instrument beam.  Histograms of the beam full width at half maximum (FWHM) values and beam ellipticities across the full array are shown in Figures \ref{fig:fwhm} and \ref{fig:ellip}, respectively.  These are all consistent with expectations from simulations of the optical properties across the field of view.

\subsection{Pixel differencing and atmospheric rejection}

Differential beam properties were investigated by normalizing the fit Gaussian beams to unity as a simple relative gain calibration and differencing the orthogonal polarization beam maps.  For {\sc polarbear}, the dominant contribution to the differential beam-induced systematic errors arises from differential pointing, or the beam center differences.  The differential pointing across the array was found to be 4.6 $\pm$ 3.0 arcseconds. 

\begin{figure}[htbp]
\begin{center}
\includegraphics[width=3.in]{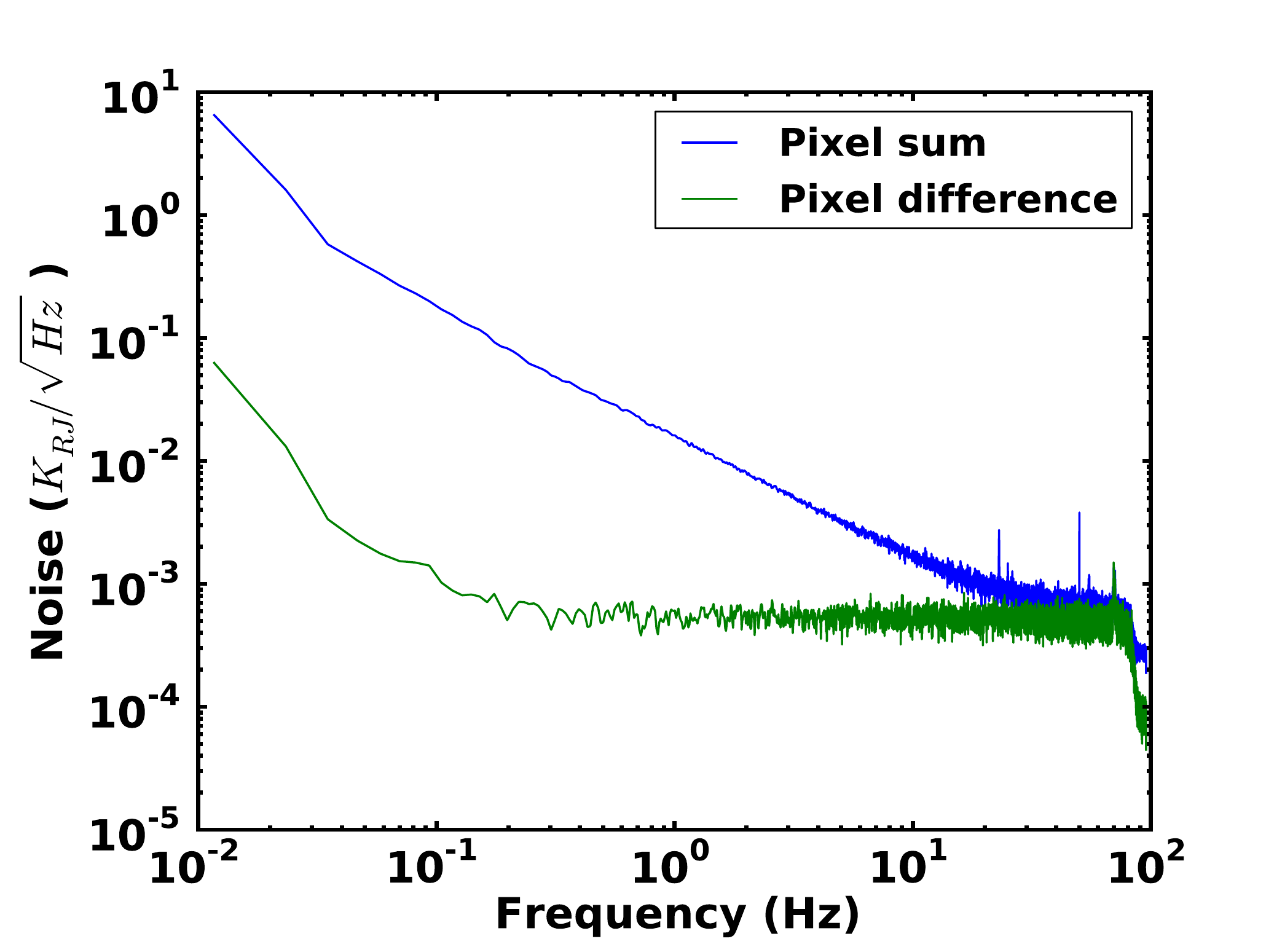}
\caption{Pixel sum and difference timestream noise spectral densities.  A simple time-variable relative calibration is made using the response to atmospheric variations in each detector.  Differencing with this simple calibration method effectively suppresses the atmospheric fluctuations over a large bandwidth.}
\label{fig:sumdiff}
\end{center}
\end{figure}

Applying a relative gain calibration that is allowed to vary with time using the atmospheric variation observed, we can find the differential timestream needed to measure the Q or U stokes parameters in individual pixels.  Fourier transforming this differential timestream allows us to see how well the unpolarized atmosphere is suppressed at low frequencies.  The $1/f$ knee of this low frequency noise informs our scan strategy to measure the large angular scale polarization signals.  Figure \ref{fig:sumdiff} shows the sum and difference amplitude spectral densities for an observation of one of our CMB patches.  Pixel differencing with this simple calibration method effectively suppresses the atmospheric fluctuations over a large bandwidth.       

\subsection{Fractional throughput and noise}

Measurements of the product of the integrated bandwidth and fractional throughput, $\eta\Delta\nu$ were also made from the beam maps presented in Section \ref{sec:beams}, as well as from earlier test of the receiver in the lab and from elevation nods of the telescope.  The fractional throughput is a measure of the percentage of the power seen by a detector from a source at the input of the receiver compared to what would be seen if the detector had perfect efficiency to that same source.  Table \ref{tab:etabw} shows the expected efficiencies of the focal plane, aperture strop, and the lenses and filters of the receiver.  Lab measurements of $\eta\Delta\nu$ using beam-filling cold loads are consistent with expectations of $\eta$ = 37\% given our measured 37 GHz integrated beamwidths.  Measurements of $\eta\Delta\nu$ can also be made using planet maps and elevation nods.  Initial results from these measurements remain consistent with $\eta = 37\%$ at the receiver window.

\begin{table}[!h]
\begin{center}
\begin{tabular}{| c |   >{\centering\arraybackslash}m{2.5cm} |}
\hline
Element & Efficiency\\
\hline
Focal plane (microwave structures and lenslets)  & 67\%  \\
Aperture stop & 87\%\\
Lenses and filters & 64\%\\
\hline
\textbf{Cumulative fractional throughput} & \textbf{37\%} \\
\hline
\end{tabular} 
\end{center}
\caption{Table of fractional throughput contribution estimates for the {\sc polarbear} receiver.  Measurements are consistent with a total receiver fractional throughput of 37\%.  }
\label{tab:etabw}
\end{table}

The design bolometer noise equivalent temperatures (NET) are about 500 $\mu K \sqrt{s}$, with variation expected due to bolometer saturation powers and atmospheric conditions. Measurements of the detector NETs can be made by similarly using beam maps or elevation nods for an absolute detector temperature gain and making a comparison to the measured noise.  Preliminary measurements from both beam maps with planets and elevation nods show a peak in the NET distribution of 550 $\mu K \sqrt{s}$.

\subsection{Polarized maps of TauA}

Maps of Tau A, a supernova remnant at the heart of the Crab nebula have been made from observtions in Chile.  Tau A is polarized by synchrotron emission and can be used by {\sc polarbear} as an astrophysical calibrator of detector polarization angles.  Regular observations of Tau A will be made at several HWP rotation angles to both characterize systematic errors and verify the detector polarization angles on the sky.  

\begin{figure}[htbp]
\begin{center}
\includegraphics[width=\textwidth]{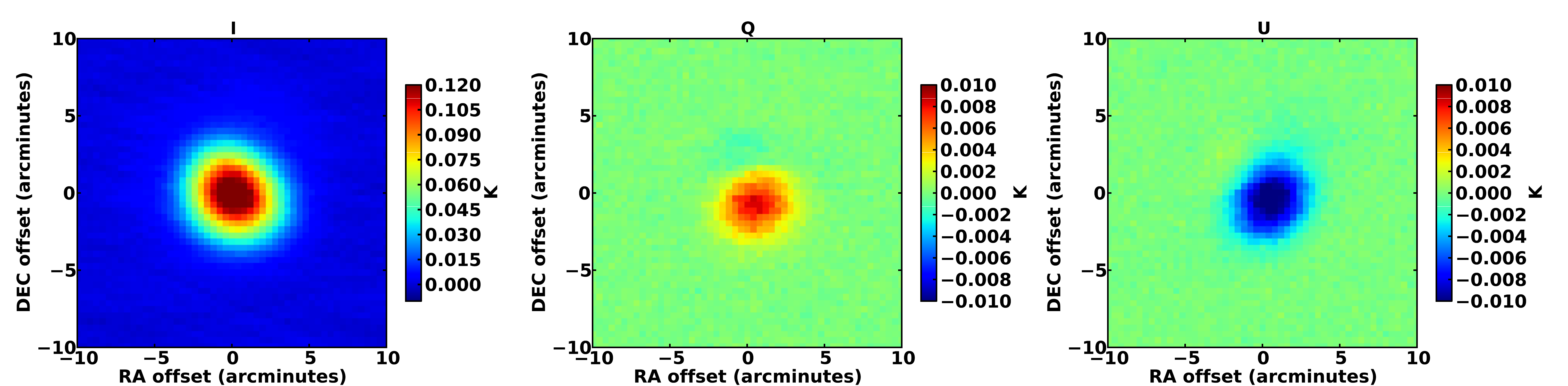}
\caption{Maps of I, Q, and U Stokes parameters from observations of the supernova remnant TauA.}
\label{fig:tauA_IQU}
\end{center}
\end{figure}

\begin{figure}[htbp]
\begin{center}
\includegraphics[width=3.in]{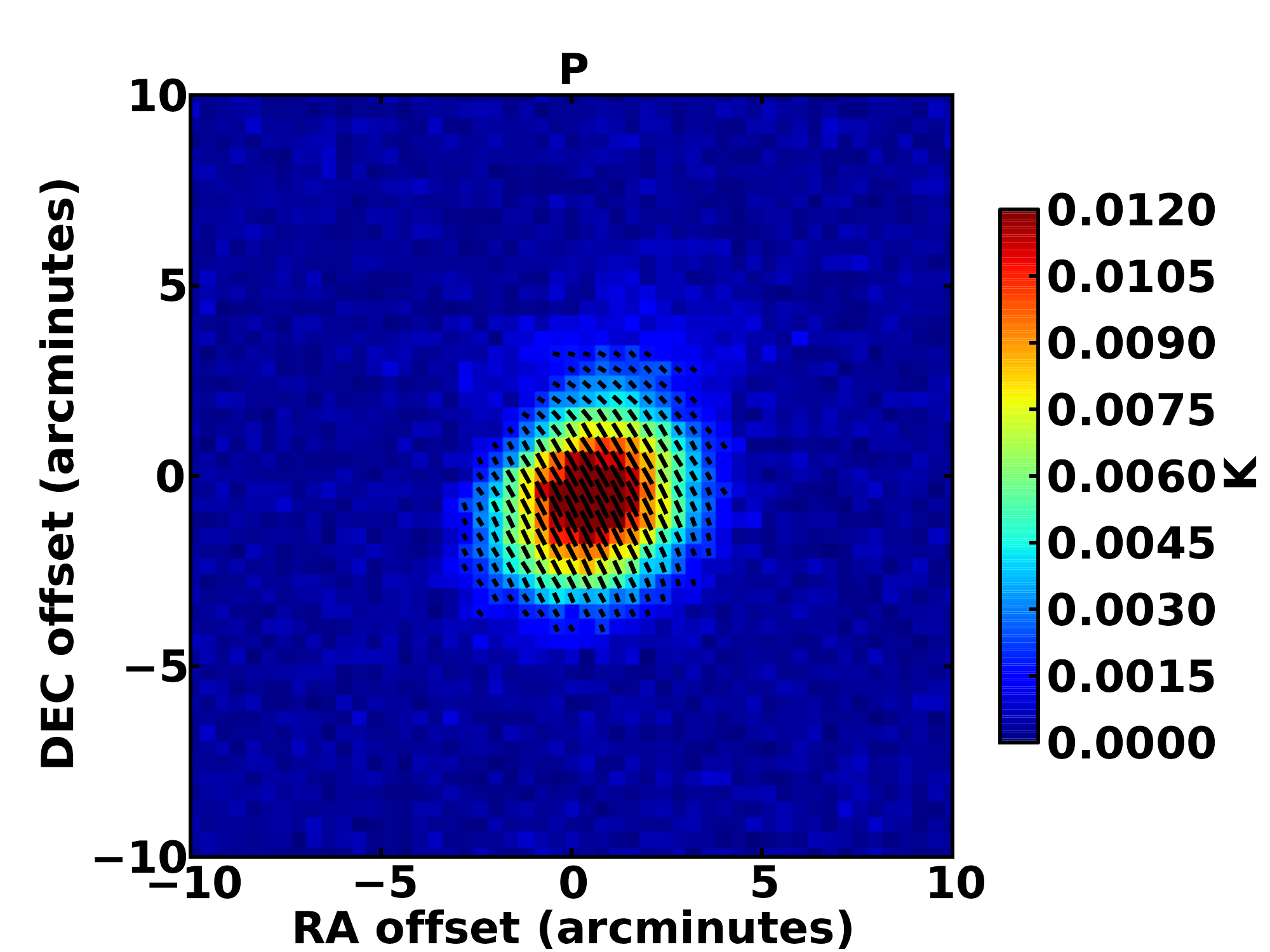}
\caption{The polarization, $P = \sqrt{Q+U}$ and polarization angle of the supernova remnant TauA.  Regular observations of TauA will be used to calibrate the detector polarization angles and to study systematic errors of the stepped HWP.}
\label{fig:tauA_P}
\end{center}
\end{figure}

Figure \ref{fig:tauA_IQU} shows the resulting I, Q, and U Stokes parameter maps of the source coadded from observations with several HWP orientations.  Studies of systematic errors related to the HWP, such as differential reflection, can be made using the maps at each HWP orientation.  Figure \ref{fig:tauA_P} shows the resulting polarization $P = \sqrt{Q+U}$ and polarization angle of the source.  

\subsection{Preliminary observations}

The instrument has been carrying out routine observations since late April of 2012.  The telescope runs on a 36 hour cycle, with $\sim$20 hours currently being used for observations of CMB patches, 4.5 hours to cycle our milliKelvin fridge, and the remaining 11.5 hours currently dedicated to calibration and instrument characterization measurements.

\begin{figure}[htbp]
\begin{center}
\includegraphics[width = 3in]{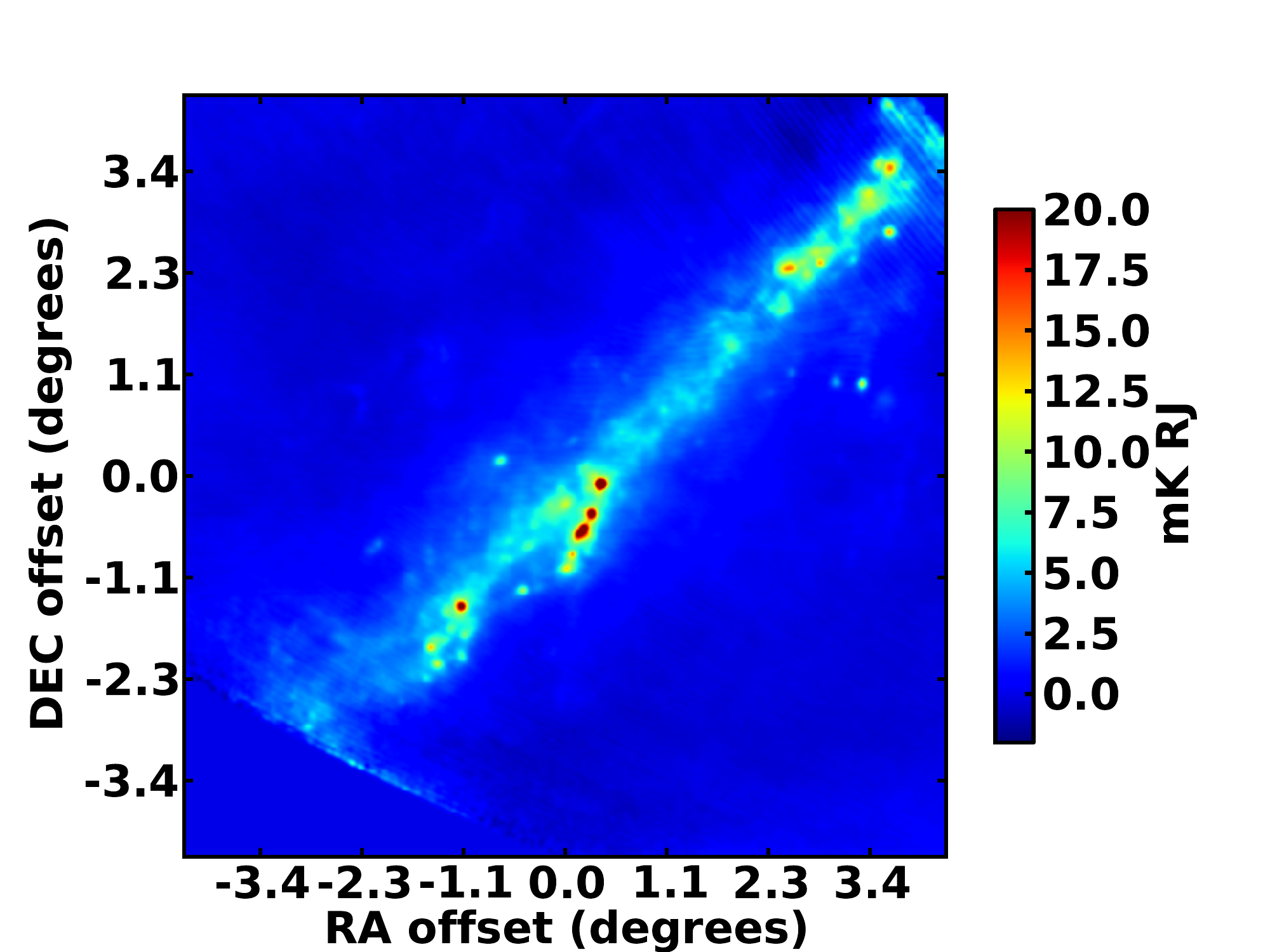}
\caption{A map of a bright region of the galaxy.}
\label{fig:pbgalbright}
\end{center}
\end{figure}

A preliminary temperature map of a patch of the galaxy with bright compact sources is shown in Figure \ref{fig:pbgalbright}.  This map, made as the $7^\circ \times 7^\circ$ patch moved across the sky, demonstrates the first order functionality of the instrument and several key analysis tasks.  Relative calibration of all detectors is carried out using a small chopped source fed through a waveguide opening in the secondary mirror.  The calibration measurements are done at the start of every constant elevation scan (CES).  A pointing model generated from many observations of point sources is applied.  Beam centers found using observations of Saturn are used to offset and co-add hundreds of individual detectors.  The features in the map due to both bright compact sources and diffuce structure in the galaxy indicate the relative success of these early analyses from the 3.5\arcmin\, resolution {\sc polarbear} instrument.  

\section{Conclusions}

The {\sc polarbear} instrument has been successfully integrated at the James Ax Observatory in the Atacama desert of Chile.  {\sc polarbear} is the first instrument to utilize a lenslet-coupled planar array detector architecture for CMB observations.  This unique focal plane technology, coupled with the Huan Tran Telescope, give {\sc polarbear} the sensitivity and angular resolution to make a detection of the B-mode polarization due to gravitational lensing on small angular scales, while still searching for the B-mode polarization on large angular scales due to inflationary gravitational waves.  Preliminary characterization of the instrument has been carried out and routine observations are now underway.

{\sc polarbear} will also serve as a unique testbed for future technologies.  The fully lithographed antenna-coupled detector architecture can be improved upon with multi-chroic antenna technologies \cite{toki_proceedings}.  Plans for future instruments utilizing these new technologies, both from the ground and in space, will be built upon the experience of developing and deploying the {\sc polarbear} instrument.    


\acknowledgments     
 
The {\sc polarbear} project is funded by the National Science Foundation under grant AST-0618398. Antenna- coupled bolometer development at Berkeley is funded by NASA under grant NNG06GJ08G. The McGill authors acknowledge funding from the Natural Sciences and Engineering Research Council and Canadian Institute for Advanced Research. MD acknowledges support from an Alfred P. Sloan Research Fellowship and Canada Re- search Chair program.  The KEK authors were supported by MEXT KAKENHI Grant Number 21111002.  All silicon wafer-based technology is fabricated at the UC Berkeley Microlab.

\bibliography{myrefs}   
\bibliographystyle{spiebib}   

\end{document}